\documentclass[preprint,showpacs,preprintnumbers,amsmath,amssymb]{revtex4}

\usepackage{epsfig}
\usepackage{dcolumn}
\usepackage{bm}
\usepackage{hyperref}
\begin{document}


\title{Colloids in a periodic potential: driven lattice gas in continuous space}


\author{Fabricio Q. Potiguar}
\email{potiguar@fisica.ufmg.br}
\author{Ronald Dickman}%
 \email{dickman@fisica.ufmg.br}
\affiliation{%
Departamento de F\'\i sica, ICEx, Universidade Federal de Minas Gerais, 30123-970,\\
Belo Horizonte, Minas Gerais, Brazil
}%



\begin{abstract}
Motivated by recent studies of colloidal particles in optical
tweezer arrays, we study a two-dimensional model of a colloidal
suspension in a periodic potential.  The particles tend to stay near
potential minima, approximating a lattice gas. The interparticle
interaction, a sum of Yukawa terms, features short-range repulsion
and attraction at somewhat larger separations, such that two
particles cannot occupy the same potential well, but occupation of
adjacent cells is energetically favored.  Monte Carlo simulation
reveals that the equilibrium system exhibits condensation, as in the
Ising model/lattice gas with conserved magnetization; the transition
appears to be continuous at a half occupancy. We study the effect of
biased hopping probabilities favoring motion along one lattice
direction, as might be generated by a steady flow relative to the
potential array. This system is found to exhibit features of the
driven lattice gas: the interface is oriented along the drive, and
appears to be smooth. A weak drive facilitates ordering of the
particles into high- and low-density regions, while stronger bias
tends to destroy order, and leads to very large energy fluctuations.
Our results suggest possible realizations of equilibrium and driven
lattice gases in a colloidal suspension subject to an optical
tweezer array.
\end{abstract}

\maketitle

\section{\label{sec1}Introduction}

Lattice gas models are an essential tool of statistical mechanics,
and have found considerable application in the study of phase
transitions in adlayers \cite{dash,schick,elgin}. In the simplest
version, particles interact only with their nearest neighbors;
equilibrium properties of the model are well understood
\cite{plishcke}. The lattice gas is readily extended to the study of
time-dependent phenomena by defining a particle dynamics (typically
via hopping) that obeys detailed balance with respect to the
Hamiltonian.  In {\it driven} lattice gas models, also known as {\it
driven diffusive systems} (DDS), \cite{zia,zia98,dicmar} particles
execute biased hopping along one of the principal lattice
directions, with periodic boundaries \cite{kat83,leu89}, leading to
a a nonequilibrium stationary state (NESS) with a nonzero current.
DDS have been studied extensively as a prototype of nonequilibrium
phase transitions. Despite their simplicity, DDS exhibit surprising
characteristics \cite{zia98} for example the {\it increase} of the
critical temperature with drive strength in the case of attractive
interactions. A driven fluid in continuous space has also been found
to possess an unusual phase diagram \cite{min05}.  In this case, the
drive appears to oppose ordering, as reflected in a reduced critical
temperature, compared with equilibrium.

Although much effort has been devoted to studying DDS theoretically
and in simulations, there are as yet no experimental realizations.
Driven colloidal suspensions offer some promise in this regard. With
the availability of optical-tweezer arrays, the dynamics of
colloidal particles in an external potential has been investigated
intensively \cite{kor02,don03,san04,chi05,lac05}.   A study of
colloidal particles in a periodic potential revealed a variety of
dynamic regimes as a function of the viscous friction coefficient
\cite{san04}.  At low friction, the motion at long times consists of
jumps between adjacent potential minima, resembling that of
particles in a lattice gas with nearest-neighbor hopping.

Motivated by this correspondence, we study a two-dimensional model
system in continuous space, of colloid-like particles in a periodic
background potential. The potential is taken so that the particles
spend most of the time near a potential minimum. The interparticle
potential, a sum of Yukawa terms, features short-range repulsion and
attraction at somewhat larger separations, such that two particles
cannot occupy the same potential well, but occupation of adjacent
cells is energetically favored.  We study the phase behavior of the
model in Monte Carlo simulations, beginning with the undriven
(equilibrium) case, followed by an exploration of the effects of
driving, and a brief examination of another nonequilibrium version,
in which there is no drive, but the external potential is time
dependent.

Although the present model includes, in the interests of
computational efficiency, certain unrealistic features, we believe
that it represents a significant step toward devising a system
capable of experimental realization, and exhibiting properties
characteristic of DDS.  Our results suggest that it is possible to
realize lattice-gas like systems, both equilibrium and driven, in a
colloidal suspension subject to a periodic external potential.

In the following section we define the model and simulation method.
In Sec. III we present simulation results, while Sec. IV contains a
summary and discussion of open issues, regarding both the model
system and possible realizations of DDS in experiments on colloids.

\section{\label{sim}Model and simulations}

We study a two-dimensional system of $N$ particles interacting via a
pairwise potential $(r)$, and subject to a periodic external
potential $V(x,y)$. The latter is of the form used to model an
optical tweezer array \cite{lac05},

\begin{equation}
\label{ppel} V(x,y)= \frac{V_0}{1+e^{-g(x,y)}},
\end{equation}
\vspace{.5em}

where $g(x,y)=A\left[\cos(2\pi x)+\cos(2\pi y)-2B\right]$. For $V_0
> 0$ the potential minima fall at the sites of the integer square
lattice $(n+1/2,m+1/2)$; parameters $A$ and $B$ together control the
well depth and the curvature near the potential minimum.  The
interparticle potential is given by the sum of a short-range
repulsive term and a slightly longer range attractive term, both of
Yukawa form \cite{dhont}:

\begin{equation}
\label{yuka} u (r) = V_1 \frac{e^{-\kappa_1 r}}{r}- V_2
\frac{e^{-\kappa_2r}}{r} - V_c,
\end{equation}
\vspace{.5em}

for $r< r_c$, and $u = 0$ for $r>r_c$. Here $\kappa_{1,2}$ are
characteristic lengths and $V_c$ is a constant taken so that $u(r_c)
= 0$. The parameter values used in this study ($\kappa_1 = 3.30$,
$\kappa_2 = 2.21$, $V_1=200$, $V_2=90$, $V_0=40$, $A=5$, $B=-0.5$,
and $r_c = 2.5$), lead to a strong short-range repulsion that
effectively prohibits two particles from occupying the same
potential well at the temperatures of interest.  The minimum of the
interparticle potential falls at $r=1$, favoring occupation of
neighboring cells. (For $r=1$, $\sqrt{2}$ and 2, one has $u=-2.347$,
-1.35 and -0.283, respectively.)  The ground state energy per
particle is -8.396, about 4.2 times that of the nearest-neighbor
square lattice gas with unit interaction.  A crude estimate of the
critical temperature is then 4.2 times that of the nearest-neighbor
lattice gas ($T_c \simeq 0.5673$), that is, $T_c \simeq 2.4$.

We perform Metropolis Monte Carlo (MC) simulations of the system
defined above. The MC time step is defined as one attempted move per
particle.  In each move, a randomly chosen particle is subject to a
random displacement $\Delta {\bf r} = (\Delta x, \Delta y)$, with
components distributed uniformly on the interval $[-1,1]$.

As in studies of DDS, the drive takes the form of a force {\bf f},
such that the work done on a particle when it suffers a displacement
$\Delta {\bf r}$ is ${\bf f} \cdot \Delta {\bf r}$.  As noted, the
system is periodic in the direction of the drive (in practice, the
$x$ direction), so that ${\bf f}$ cannot be written as the gradient
of a potential. The acceptance probability for a particle
displacement $\Delta {\bf r}$ is:

\begin{equation}
\label{met-step}
 P(\Delta {\bf r})= \min \left\{1, \exp
\left[-\beta(\Delta E-{\bf f}\cdot \Delta {\bf r})\right] \right\},
\end{equation}
\vspace{.5em}

where $\Delta E$ is the change in energy and $\beta$ represents
inverse temperature, in units such that $k_B=1$.

The quantities of principal interest are the average energy $\langle
E \rangle$, the order parameter ${\bf m}= (m_x, m_y)$, and the
corresponding fluctuations. To define the order parameter we
introduce lattice-gas variables: $\sigma_{ij} = 1$ if the cell
centered at $(i+1/2,j+1/2)$ is occupied, and is zero otherwise. Then
an appropriate order parameter for a lattice gas with conserved
particle density can be defined, for a system of $L \times L$ sites,
via \cite{dicmar}

\begin{equation}
\label{op-comp} m_x =
\frac{1}{L^3}\sum_{j=1}^{L}\left[\sum_{i=1}^{L}(1-2\sigma_{ij})\right]^2,
\end{equation}
\vspace{.5em}

with $m_y$ given by exchanging indices $i$ and $j$ in the sums.  In
a disordered phase $\langle m_x \rangle = \langle m_x \rangle = 0$.
In a half-occupied system, a maximally ordered, isotropic
configuration (a square of side $L/\sqrt{2}$) has $ m_x  =  m_y  =
1/2$.  In a system with periodic boundaries, however, the surface
energy is smaller in a strip configuration.  If the particles occupy
all sites the region $j_0 \leq j \leq j_0 + L/2$, then $m_x = 1$ and
$m_y = 0$; the values are exchanged for a strip oriented along the
$y$ direction.  $m = \sqrt{m_x^2 + m_y^2}$ characterizes overall
ordering.  In equilibrium, the two orientations are equally likely,
but for low temperatures and reasonably large systems, the typical
time for switching between them is much larger than practical
simulation times, so it is interesting to characterize the degree of
order by the anisotropy parameter $\Delta m \equiv m_> - m_<$, where
$m_> \equiv \max \{\langle m_x \rangle, \langle m_y \rangle \}$, and
$m_<$ denotes the lesser of the two components. Experience with DDS
shows that in driven systems $m_>$ is always along the drive
direction.

\section{\label{res}Numerical results}

\subsection{Equilibrium properties}

\begin{figure}[h]
\rotatebox{0}{\epsfig{file=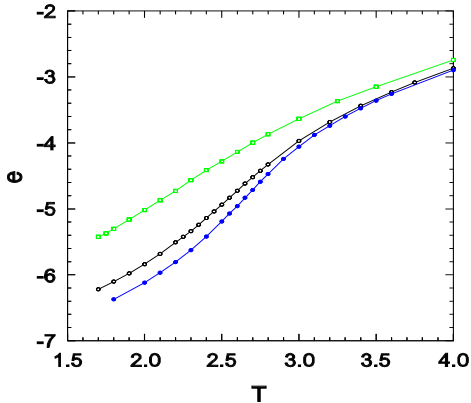,width=7.0cm,height=7.0cm}}
\caption{Energy per particle versus temperature for $L= 10$, 20 and
30 (upper to lower). \label{colep}}
\end{figure}

We study half-occupied systems, $N = L^2/2$, in periodic cells of
linear size $L=10$, 20 and 30.  Our results represent averages over
the stationary regime of 10 or more independent realizations, each
of $10^6$ or more time steps.  In Fig. \ref{colep} we show the mean
energy $e$ per particle, which exhibits the qualitative behavior
typical of lattice gases or fluids with a hard core and short-range
attraction. Evidence of a phase transition is seen in the
progressive growth, with system size, of the peak in the specific
heat per particle, $c$ = var($E$)/$NT^2$, shown in Fig. 2.  The
specific heat maxima fall at temperatures $T_m=2.1$, 2.5 and 2.63
for sizes $L=10$, 20 and 30, respectively. These data are fit very
well by the expression $T_m = T_c + Const./L$, yielding the estimate
$T_c = 2.90(1)$ for the transition temperature.

\begin{figure}[h]
\rotatebox{0}{\epsfig{file=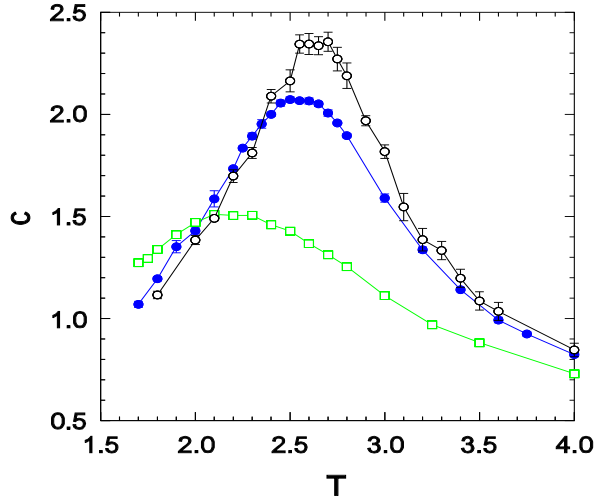,width=7.0cm,height=7.0cm}}
\caption{Specific heat per particle versus temperature for $L= 10$,
20 and 30, in ascending order. \label{colve}}
\end{figure}

In Fig. 3 we show the order parameter $m$ and the anisotropy $\Delta
m$ for $L=30$.  These results are consistent with a transition near
$T=2.9$.  Of note is the much smaller finite size effect in $\Delta
m$ as compared with $m$ in the high-temperature regime, as the
fluctuations in the two components of {\bf m} are nearly equal. The
inset shows that $\chi \equiv \frac{L^2}{T} \mbox{var} (m)$ exhibits
a well defined maximum in the vicinity of the transition.

Based on the present results, we can conclude that the half-filled
system exhibits an (apparently continuous) order-disorder transition
near a temperature of 2.9.  Further details on the nature of the
transition (which from symmetry considerations should belong to the
Ising model universality class), must await more extensive studies
using larger systems.

To close this section we show a representative configuration (Fig.
4) for $L=30$ at temperature $T=2$.  Here the system has separated
into distinct high- and low-density phases.  The latter is quite
dilute, while the former clearly reflects the periodic potential,
and possesses a low density of vacant sites. The interface between
the high- and low-density regions is quite rough.

\begin{figure}[h]
\rotatebox{0}{\epsfig{file=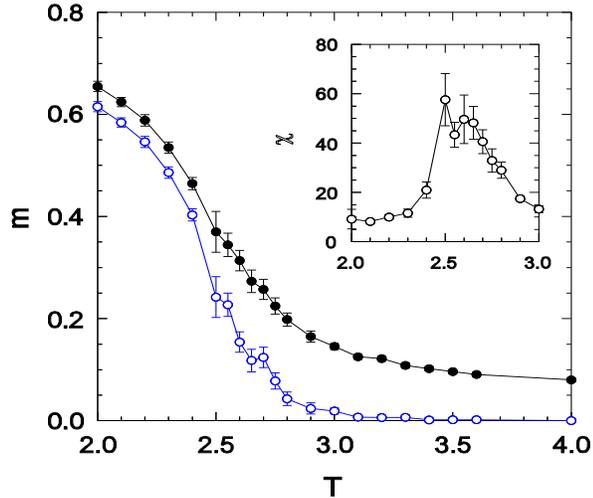,width=7.0cm,height=7.0cm}}
\caption{Magnetization $m$ (filled symbols) and anisotropy $\Delta
m$ (open symbols) versus temperature for $L= 30$. Inset: scaled
variance of the order parameter. \label{m30}}
\end{figure}

\begin{figure}[h]
\rotatebox{0}{\epsfig{file=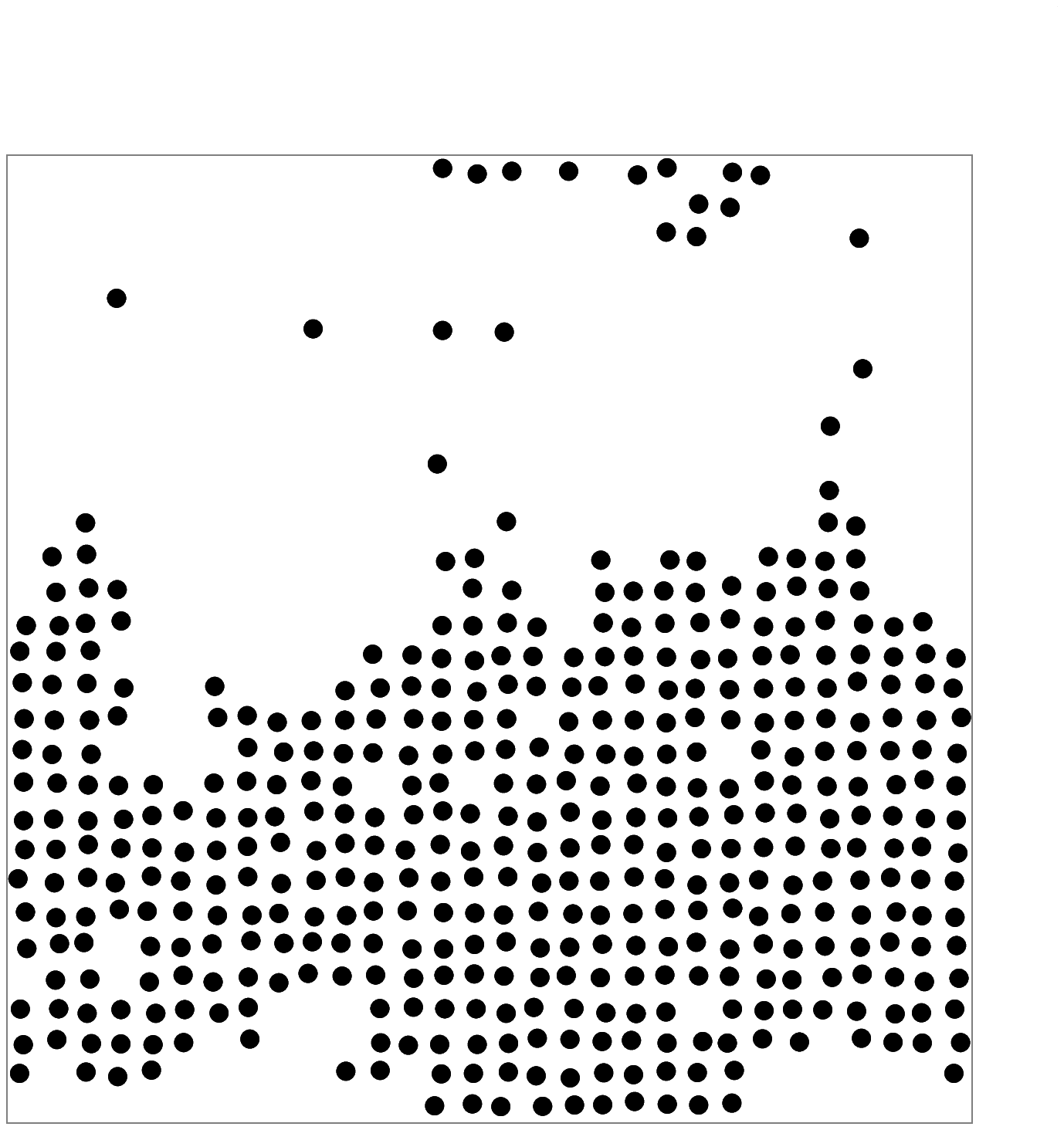,width=7.0cm,height=7.0cm}}
\caption{Typical configuration, $L= 30$, $T=2$. \label{cfndt2}}
\end{figure}

\subsection{Driven system}

\begin{figure}[h]
\rotatebox{0}{\epsfig{file=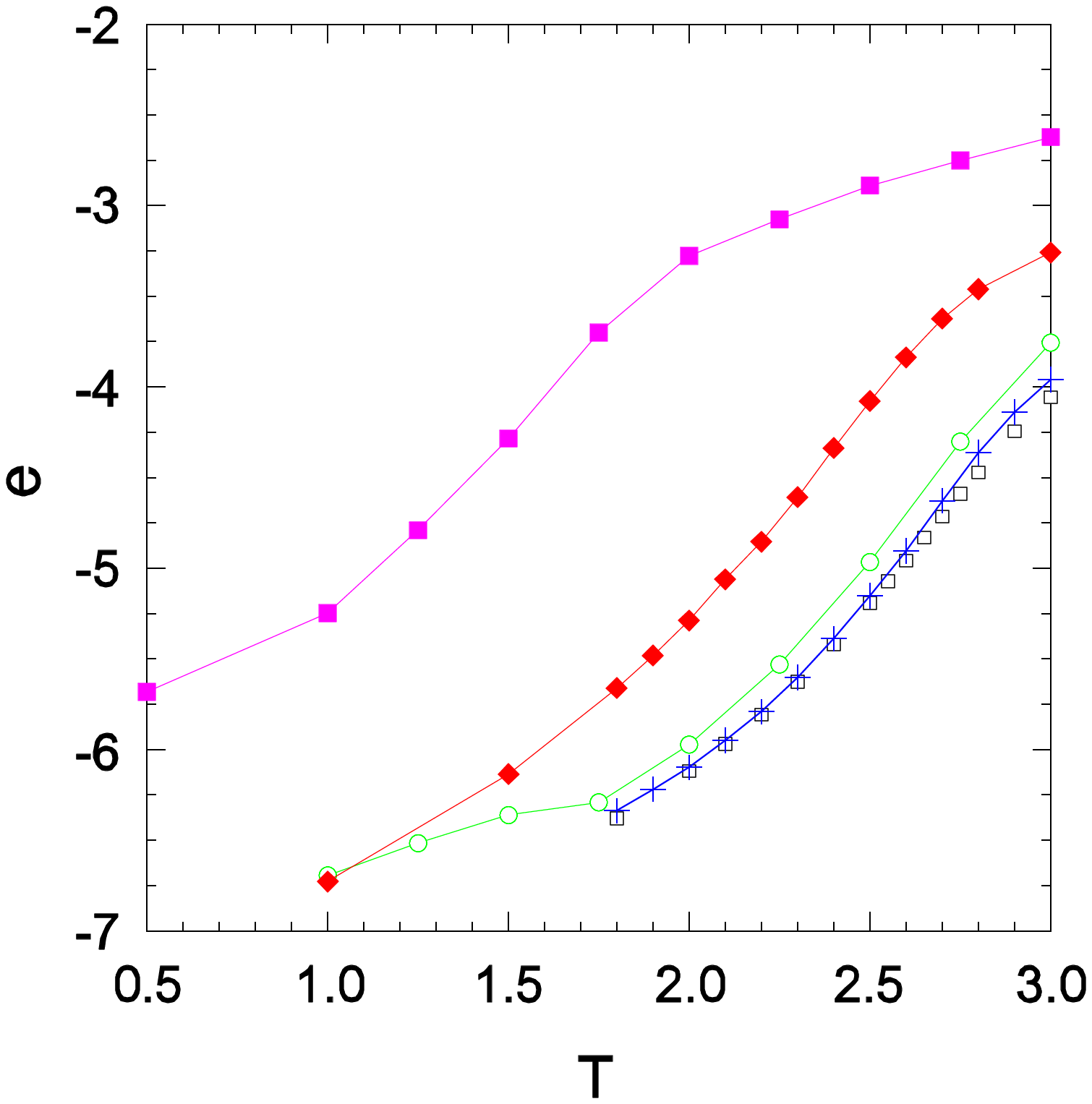,width=7.0cm,height=7.0cm}}
\caption{Mean energy per particle versus temperature for $L=30$ and
(right to left) drive strength $f=0$, 1, 2, 4 and 7.} \label{ep30}
\end{figure}

We determined the stationary properties for various drive strengths
$f$, for system sizes $L=20$ and $L=30$, following the procedure
described above. The mean energy is plotted versus temperature in
Fig. \ref{ep30} for various drive intensities. For the relatively
weak drive ($f=1$), the energy is just slightly higher than in
equilibrium. For the larger drives we observe a substantial increase
in energy.

\begin{figure}[h]
\rotatebox{0}{\epsfig{file=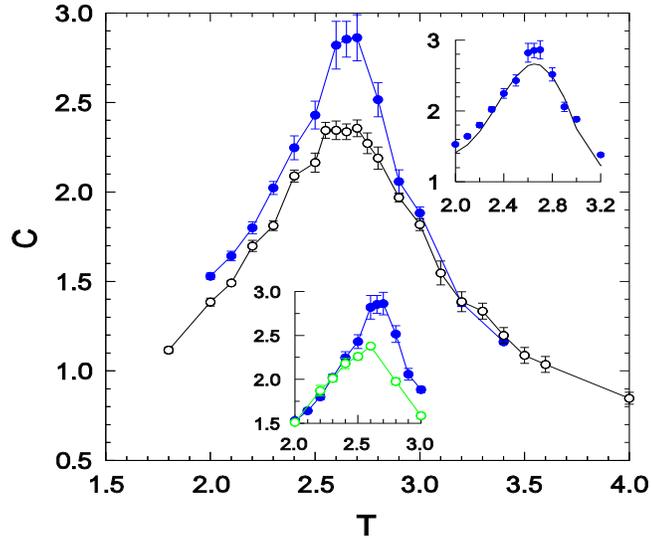,width=7.0cm,height=7.0cm}}
\caption{$c \equiv $var$(E)/NT^2$ versus $T$ for drive $f=1$ (upper
curve) and in equilibrium (lower).  Upper inset: comparison, for
$f=1$, of the scaled variance $c$ (points) and $de/dT$ (smooth
curve). Lower inset: scaled variance for $f=1$ and $L=20$ (open
symbols) and $L=30$ (filled symbols).} \label{vare01}
\end{figure}

In Fig. \ref{vare01} we compare the scaled energy variance $c \equiv
$var$(E)/NT^2$ in equilibrium and under drive $f=1$.  Although this
quantity does not represent the specific heat for $f>0$, it is
nevertheless reasonable to suppose that a singularity in $c$ (or a
sharp peak, in a finite system), marks a phase transition.  It is
therefore interesting to note that the $f=1$ data exhibit a sharper
peak (and at a slightly higher temperature), than in equilibrium.
This suggests that a weak drive facilitates ordering. For $f=1$, the
scaled variance (upper inset) follows the same trends as, but is
generally greater than, the specific heat, $de/dT$. (In this case we
obtain $de/dT$ via numerical differentiation of a polynomial fit to
the energy data.) This suggests that the drive causes fluctuations
beyond those generated by thermal mechanisms.  (That is, the drive
realizes work on the system but is not included in the energy
balance.) The dependence of var$(e)$ on system size (lower inset) is
qualitatively similar to that found in equilibrium; we estimate $T_c
= 2.85(5)$ for $f=1$.

\begin{figure}[h]
\rotatebox{0}{\epsfig{file=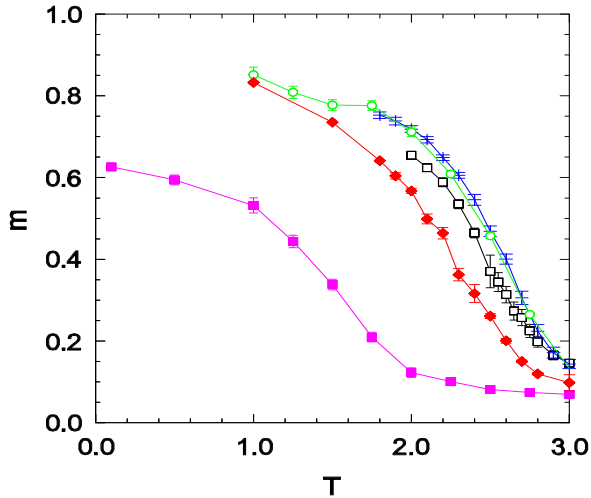,width=7.0cm,height=7.0cm}}
\caption{Order parameter versus temperature for $L=30$ and drive
$f=0$ (open squares), $f=1$ (+), $f=2$ (open circles), $f=4$
(diamonds), and $f=7$ (filled squares).} \label{m30dr}
\end{figure}

The plot of the order parameter (Fig. \ref{m30dr}) confirms that a
weak drive ($f=1$, 2) enhances ordering, whereas a stronger one
inhibits it. These data suggest a transition temperature of $T_c
\approx 2$, for $f=7$.

\begin{figure}[h]
\rotatebox{0}{\epsfig{file=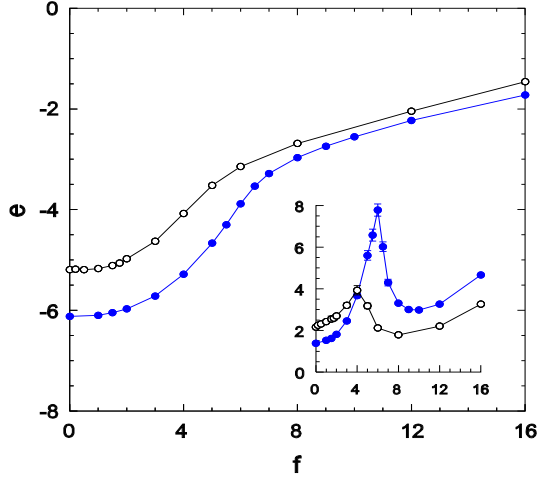,width=7.0cm,height=7.0cm}}
\caption{Energy per particle versus drive $f$ for temperature
$T=2.5$ (open symbols), and $T=2.0$ (filled symbols), $L=30$. Inset:
$c = $var$(E)/NT^2$ for the same parameters.} \label{evfa}
\end{figure}

\begin{figure}[h]
\rotatebox{0}{\epsfig{file=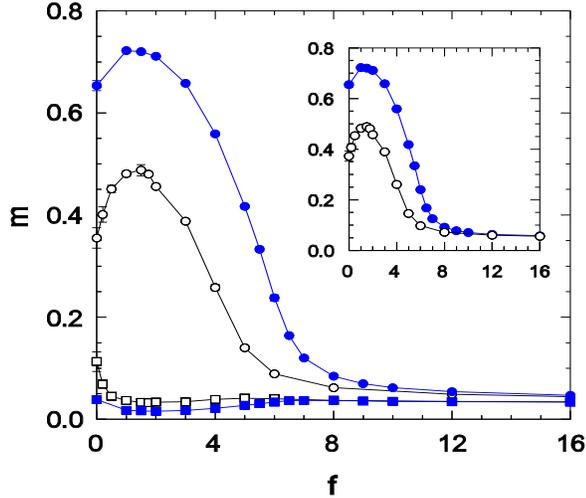,width=7.0cm,height=7.0cm}}
\caption{Order parameter components $m_>$ (upper curves) and $m_<$
(lower curves) versus drive $f$ for $T=2.5$ (open symbols), and
$T=2.0$ (filled symbols), $L=30$.  Inset: $m = \sqrt{m_>^2 + m_<^2}$
versus $f$.} \label{mvfa}
\end{figure}

The effect of increasing drive at fixed temperature is shown in
Figs. \ref{evfa} (energy and its variance) and \ref{mvfa} (order
parameter). The energy increases slowly with $f$ for a weak drive,
rapidly for intermediate drive strength, and then exhibits a steady,
more gradual growth for $f > 8$ or so. The energy variance shows a
marked peak (near $f=4$ for $T=2.5$, and $f=6$ for $T=2.0$), which
appears to be associated with destruction of an ordered arrangement.
The amplitude of this peak is much larger than in equilibrium.  For
larger drives, var$(e)$ increases steadily with $f$, as the drive
forces particles out of the periodic potential minima. Fig.
\ref{mvfa} again shows that a weak drive enhances ordering. The
order parameter reaches a maximum near $f=1$ - 2, and then decays
rapidly when the drive is increased further, and the driving force
begins to dominate interparticle attraction.

\begin{figure}[h]
\rotatebox{0}{\epsfig{file=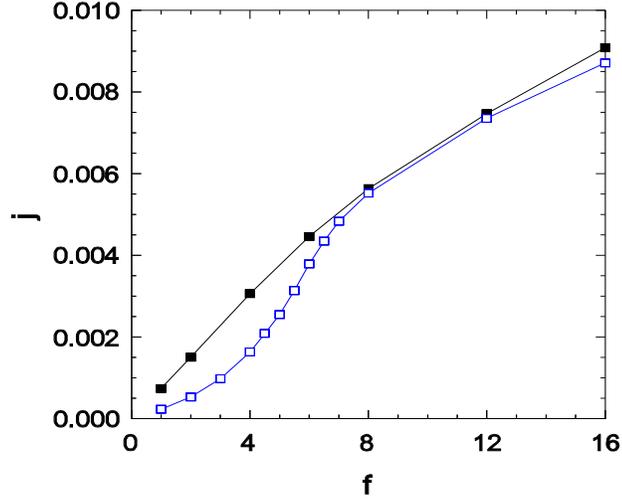,width=7.0cm,height=7.0cm}}
\caption{Current density $j$ versus drive $f$ for $T=3.5$ (filled
symbols), and $T=2.0$ (open symbols), $L=30$.} \label{jt35}
\end{figure}

The stationary current density $j$, defined as the mean displacement
$\langle \Delta z \rangle$ per site and unit time, is plotted as a
function of drive in Fig. \ref{jt35}.  For the range of parameters
studied here, $j$ is an increasing function of both $f$ and $T$. In
the disordered phase (upper set of points in Fig. \ref{jt35}), the
current grows linearly with $f$ for small $f$, and then shows signs
of saturating at larger values of the drive. At lower temperatures
and weak drives, such that the system is ordered, the current is
severely reduced, but it takes values comparable to those at higher
temperature once $f$ is large enough to disorder the system.  The
latter event is signaled by a sharp peak in the variance of the
energy (Fig. \ref{evfa}); the singularity (if any) in the current is
much weaker.

\begin{figure}[h]
\rotatebox{0}{\epsfig{file=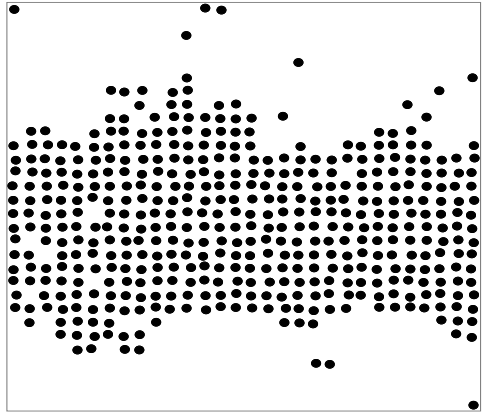,width=7.0cm,height=7.0cm}}
\caption{Typical configuration, $L=30$, $f=1.0$, $T=2.0$. The drive
is directed to the right. \label{conf-1}} \vspace{1em}
\end{figure}

\begin{figure}[h]
\rotatebox{0}{\epsfig{file=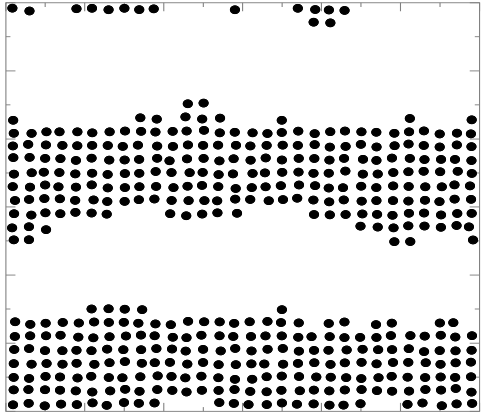,width=7.0cm,height=7.0cm}}
\caption{Typical configuration, $L=30$, $f=1.0$, $T=1.1$.
\label{conf-2}}
\end{figure}

Our results suggest that the particles are organized into dense
stripes along the field direction, as in the driven lattice gas. The
configurations shown in Figs. \ref{conf-1} ($T=2$, $f=1$), and
\ref{conf-2} ($T=1.1$, $f=1$), confirm this expectation.  Comparing
Figs. \ref{cfndt2} ($f=0$) and \ref{conf-1} ($f=0$), the chief
difference is the smoother interface found in the driven system.
Fig. \ref{conf-2} shows a configuration well below the transition
temperature, in which two separate stripes have emerged. As in the
driven lattice gas, this appears to be a long-lived metastable state
at low temperature \cite{dicmar}.

$\;$ \vspace{2cm}

\subsection{Moving background potential}

We performed some preliminary studies of another method for
perturbing the system out of equilibrium.  In this case there is no
drive, but the external potential is time dependent, given by
$V(x-vt,y)$, with the function $V(x,y)$ as in Eq. (\ref{ppel}).
[Since the potential is periodic, the first argument of $V$ is
effectively $x - vt \;(\mbox{mod } 1)$.]  The particles are dragged
along by the moving potential array, as in {\it optical peristalsis}
\cite{koss}.

\begin{figure}[h]
\rotatebox{0}{\epsfig{file=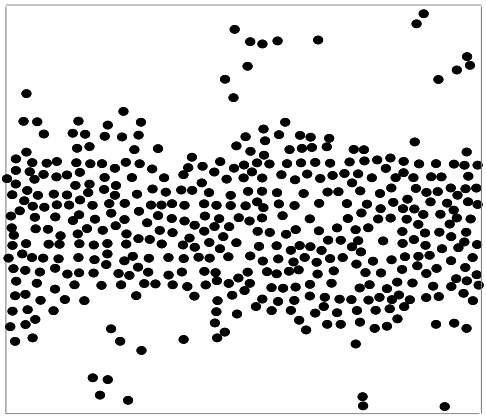,width=7.0cm,height=7.0cm}}
\caption{Typical configuration under moving background potential,
$L=30$, $v=0.1$, $T=1.5$. \label{v1t15}}
\end{figure}

In these studies we use $L=30$; the background potential amplitude
$V_0$ is reduced from 40 to 10, and the trial particle displacement
$\Delta x$ is uniform on $[-.5,.5]$ (similarly for $\Delta y$).
Otherwise the parameters are as in the studies reported above. For
velocities $v \leq 0.02$, there is anisotropic, lattice-gas like
ordering for $T < T_c$.  (The critical temperature is about 2.5 for
$v=0.01$.) For larger velocities, $v = 0.05$ or 0.1, the critical
temperature is reduced considerably ($T_c \simeq 1.8$ for $v=0.1$).
The interface between phases is again oriented along the $x$
direction, but for $v \geq 0.05$ the particle configuration (Fig.
\ref{v1t15}) shows no hint of periodic structure.

\section{\label{con}Conclusions}

We study a two dimensional model of colloidal particles subject to a
periodic external potential and an interaction that favors ordering
in a lattice-gas-like arrangement.  Stationary properties of the
model are determined via Monte Carlo simulation.  In equilibrium the
system appears to undergo a second order phase transition, as
reflected in the specific heat and the order parameter.  When the
short-range hopping dynamics is biased by a drive, we observe
anisotropic phase separation, i.e., the interface is oriented along
the drive, and becomes smoother, as found in studies of the driven
lattice gas \cite{leung88}.  A weak drive facilitates ordering, as
in the driven lattice gas.  A strong drive, by contrast, tends to
destroy order, and provokes very large energy fluctuations.
Preliminary analysis indicates that a colloidal system under a
moving background potential may offer another realization of
anisotropic phase separation.

Our results suggest that a system of particles interacting via a
potential featuring short-range repulsion and slightly longer range
attraction, in the presence of a periodic external potential with a
lattice constant matched to the interparticle potential minimum,
provides a good candidate for realization of both equilibrium and
driven lattice gases.  Although the physics of such a continuous
space system is richer than the corresponding lattice model, there
is good reason to expect that similar scaling properties will
emerge, in a suitable range of parameter values.

While the equilibrium model could in principle be realized using a
planar optical tweezer array to provide the external potential, the
driven system must, by nature, have periodic boundary conditions
along the drive direction.  This suggests a cylindrical geometry,
with the driving force provided by Stokes drag on the particles in
steady Couette flow.  (This scheme, without the optical tweezer
array, could be used to realize the type of system studied in Ref.
\cite{min05}.) Alternatively, one could impose relative rotational
motion between the colloidal suspension and the tweezer array. We
defer analysis of the feasibility of such a setup to future work.

Many additional aspects of the system remain to be explored
theoretically and in simulations.  While the present study uses
(because of its greater efficiency) Monte Carlo simulation to map
out equilibrium properties and nonequilibrium steady states, the
driven system should be investigated via Langevin dynamics. Aside
from affording a better account of nonequilibrium stationary and
time-dependent properties, this method would allow for inclusion of
hydrodynamic interactions between the particles and with container
walls.  Preliminary studies of the model system studied here using
the Langevin equation in fact yield qualitatively similar results to
those reported above \cite{inprog}. It would also be interesting to
remove the periodic boundaries in the direction perpendicular to the
drive, and the restriction to two dimensions. More detailed
characterization of the phase transitions exhibited by this system,
both in and out of equilibrium, using large-scale simulations, are
planned for future work.

\section*{Acknowledgements}

We are grateful to Oscar Nassif de Mesquita for helpful discussions.
This work is supported by CNPq and Fapemig, Brazil.


\begin{thebibliography}{100}
\bibitem{dash}
         M. Bretz and J. G. Dash,
         Phys. Rev. Lett. {\bf 27}, 647 (1971).

\bibitem{schick}
         M. Schick and R. L. Siddon,
         Phys. Rev, A {\bf 8}, 339 (1973).

\bibitem{elgin}
         R. L. Elgin and D. L. Goodstein,
         Phys. Rev. A {\bf 9}, 2657 (1974).

\bibitem{plishcke}
        M. Plischke and B. Bergensen,
        {\it Equilibrium statistical physics}
        (World Scientific Publishing Co., Singapore, 2003).

\bibitem{zia}B. Schmittmann and R. K. P. Zia, {\em Statistical mechanics
of driven diffusive systems}, vol. 17 of {\em Phase transition and
critical phenomena}, eds. C. Domb and J. L. Lebowitz (Academic Press,
London, 1995).

\bibitem{zia98}B. Schmittmann and R. K. P. Zia, Phys. Rep. {\bf 301}, 45 (1998).

\bibitem{dicmar}J. Marro, R. Dickman, {\em Nonequilibrium phase transitions
in lattice models} (Cambridge University Press, Cambridge, 1999).

\bibitem{kat83}
        S. Katz, J. L. Lebowitz, and H. Spohn,
        Phys. Rev. B {\bf 28}, 1655 (1983).

\bibitem{leu89}
        K.-t. Leung, B. Schmittmann, and R. K. P. Zia {\bf 62}, 1772 (1989).

\bibitem{min05}
        M. D\'\i ez-Minguito, P. L. Garrido, J. Marro, Phys. Rev. E
{\bf 72}, 26103 (2005).


\bibitem{san04}J. M. Sancho, A. M. Lacasta, K. Lindenberg, I. M. Sokolov,
A. H. Romero, Phys. Rev. Lett. {\bf 92}, 250601 (2004).


\bibitem{kor02}
        P. T. Korda, M. B. Taylor, D. G. Grier,
        Phys. Rev. Lett. {\bf 89}, 128301 (2002).

\bibitem{don03}M. P. MacDonald, G. C. Spalding, K. Dholakia, Nature
{\bf 426}, 421 (2003).

\bibitem{chi05}P. Y. Chiou, A. T. Ohta, M. C. Wu, Nature {\bf 436}, 370 (2005).

\bibitem{lac05}
         A. M. Lacasta, J. M. Sancho, A. H. Romero, K. Lindenberg,
         Phys. Rev. Lett. {\bf 94}, 160601 (2005).

\bibitem{dhont}
         J. K. G. Dhont, {\em An introduction to dynamics of colloids} (Elsevier, Amsterdam, 2003).


\bibitem{koss}
         B. A Koss and D. G. Grier,
         App. Phys. Lett. {\bf 82}, 3985 (2003).

\bibitem{leung88}
         K.-t. Leung, K. K. Mon, J. L. Vall\'es, and R. K. P. Zia,
         Phys. Rev. Lett. {\bf 61}, 1744 (1988).

\bibitem{inprog}
         F. Q. Potiguar and R. Dickman, {\em in preparation}.

\end{thebibliography}
\end{document}